\documentclass[10pt]{IEEEtran}

\pdfoutput=1

\usepackage{graphicx}   
\usepackage{amssymb}
\usepackage{amsmath}
\usepackage{epstopdf} 
\usepackage{xspace}
\usepackage{cancel}

\usepackage[stretch=16,shrink=16,step=4]{microtype}

\usepackage{amssymb}
\usepackage{bbm}
\usepackage{mathrsfs}
\usepackage{xspace}

\newenvironment{textbmatrix}{	\setlength{\arraycolsep}{2.5pt}%
								\big[\begin{matrix}}{\end{matrix}\big]%
								\raisebox{0.08ex}{\vphantom{M}}}

\def\be{\begin{equation}}
\def\ee{\end{equation}}
\def\benn{\begin{equation*}}
\def\eenn{\end{equation*}}
\def\een{\nonumber \end{equation}}
\def\mat{\begin{bmatrix}}
\def\emat{\end{bmatrix}}
\def\btm{\begin{textbmatrix}}
\def\etm{\end{textbmatrix}}

\def\ba#1\ea{\begin{align}#1\end{align}}
\def\bann#1\eann{\begin{align*}#1\end{align*}}
\def\bs#1\es{\begin{split}#1\end{split}} 
\def\bg#1\eg{\begin{gather}#1\end{gather}} 
\def\bi#1\ei{\begin{itemize}#1\end{itemize}} 
\def\bsa#1\esa{\begin{IEEEeqnarray}{rCl}#1\end{IEEEeqnarray}}
\def\bsann#1\esann{\begin{IEEEeqnarray*}{rCl}#1\end{IEEEeqnarray*}}

\newcommand{\safemath}[2]{\newcommand{#1}{\ensuremath{#2}\xspace}}

\newcommand{\lefto}{\mathopen{}\left}

\DeclareMathOperator*{\argmin}{arg\;min}		% arg min
\DeclareMathOperator{\Exop}{\mathbb{E}}		% expectation operator
\safemath{\normal}{\mathcal{N}}				% normal distribution
\safemath{\circnorm}{\mathcal{CN}}			% circ. symm. normal
\safemath{\uniform}{\mathcal{U}}				% uniform distribution
\safemath{\interior}{\mathrm{Int}}			 % interior of a set
\safemath{\dfn}{:=}							% definition
\newcommand{\dotgeq}{\dot{\geq}} % dot greater than
\newcommand{\dotleq}{\dot{\leq}} % dot less than
\newtheorem{theorem}{Theorem}
\newtheorem{remark}{Remark}

\newcommand{\prob}[1]{\ensuremath{\mathbb{P}\lefto[#1\right]}}% Probability 

\safemath{\SNR}{\text{\sc snr}} 				% signal to noise ratio
\safemath{\No}{N_0}							% noise spectral density
\safemath{\Es}{E_s}							% energy per symbol
\safemath{\Eb}{E_b}							% energy per bit
\safemath{\EbNo}{\frac{\Eb}{\No}}
\safemath{\EsNo}{\frac{\Es}{\No}}

\providecommand{\Hop}{\ensuremath{\mathbb{H}}} % channel operator
\safemath{\LH}{L_{\Hop}}						% Weyl symbol
\safemath{\SH}{S_\Hop}						% spreading function
\safemath{\HH}{H_{\Hop}}						% transfer function
\safemath{\CH}{C_\Hop}						% scattering function
\safemath{\RH}{R_\Hop}						% TF correlation function
\safemath{\Rh}{R_h}							% correlation function

\safemath{\dB}{\,\mathrm{dB}}
\safemath{\dBm}{\,\mathrm{dBm}}
\safemath{\Hz}{\,\mathrm{Hz}}
\safemath{\kHz}{\,\mathrm{kHz}}
\safemath{\MHz}{\,\mathrm{MHz}}
\safemath{\GHz}{\,\mathrm{GHz}}
\safemath{\s}{\,\mathrm{s}}
\safemath{\ms}{\,\mathrm{ms}}
\safemath{\mus}{\,\mathrm{\mu s}}
\safemath{\ns}{\,\mathrm{ns}}
\safemath{\meter}{\,\mathrm{m}}
\safemath{\mm}{\,\mathrm{mm}}
\safemath{\cm}{\,\mathrm{cm}}
\safemath{\m}{\,\mathrm{m}}
\safemath{\W}{\,\mathrm{W}}
\safemath{\J}{\,\mathrm{J}}
\safemath{\K}{\,\mathrm{K}}
\safemath{\bit}{\,\mathrm{bit}}

\safemath{\define}{\triangleq}			% definition

\safemath{\equivalent}{\sim}
\safemath{\distas}{\sim}					% distributed according to

\safemath{\reals}{\mathbb{R}}
\safemath{\positivereals}{\mathbb{R}^{+}}
\safemath{\integers}{\mathbb{Z}}
\safemath{\posint}{\mathbb{Z}_{+}}
\safemath{\naturals}{\mathbb{N}}
\safemath{\complexset}{\mathbb{C}}

\safemath{\setA}{\mathcal{A}}
\safemath{\setB}{\mathcal{B}}
\safemath{\setC}{\mathcal{C}}
\safemath{\setD}{\mathcal{D}}
\safemath{\setE}{\mathcal{E}}
\safemath{\setF}{\mathcal{F}}
\safemath{\setG}{\mathcal{G}}
\safemath{\setH}{\mathcal{H}}
\safemath{\setI}{\mathcal{I}}
\safemath{\setJ}{\mathcal{J}}
\safemath{\setK}{\mathcal{K}}
\safemath{\setL}{\mathcal{L}}
\safemath{\setM}{\mathcal{M}}
\safemath{\setN}{\mathcal{N}}
\safemath{\setO}{\mathcal{O}}
\safemath{\setP}{\mathcal{P}}
\safemath{\setQ}{\mathcal{Q}}
\safemath{\setR}{\mathcal{R}}
\safemath{\setS}{\mathcal{S}}
\safemath{\setT}{\mathcal{T}}
\safemath{\setU}{\mathcal{U}}
\safemath{\setV}{\mathcal{V}}
\safemath{\setW}{\mathcal{W}}
\safemath{\setX}{\mathcal{X}}
\safemath{\setY}{\mathcal{Y}}
\safemath{\setZ}{\mathcal{Z}}
\safemath{\emptySet}{\varnothing}

\safemath{\bma}{\mathbf{a}}
\safemath{\bmb}{\mathbf{b}}
\safemath{\bmc}{\mathbf{c}}
\safemath{\bmd}{\mathbf{d}}
\safemath{\bme}{\mathbf{e}}
\safemath{\bmf}{\mathbf{f}}
\safemath{\bmg}{\mathbf{g}}
\safemath{\bmh}{\mathbf{h}}
\safemath{\bmi}{\mathbf{i}}
\safemath{\bmj}{\mathbf{j}}
\safemath{\bmk}{\mathbf{k}}
\safemath{\bml}{\mathbf{l}}
\safemath{\bmm}{\mathbf{m}}
\safemath{\bmn}{\mathbf{n}}
\safemath{\bmo}{\mathbf{o}}
\safemath{\bmp}{\mathbf{p}}
\safemath{\bmq}{\mathbf{q}}
\safemath{\bmr}{\mathbf{r}}
\safemath{\bms}{\mathbf{s}}
\safemath{\bmt}{\mathbf{t}}
\safemath{\bmu}{\mathbf{u}}
\safemath{\bmv}{\mathbf{v}}
\safemath{\bmw}{\mathbf{w}}
\safemath{\bmx}{\mathbf{x}}
\safemath{\bmy}{\mathbf{y}}
\safemath{\bmz}{\mathbf{z}}

\safemath{\bmxi}{\boldsymbol{\xi}}
\safemath{\bmlambda}{\mathbf{\lambda}}
\safemath{\bmmu}{\mathbf{\mu}}
\safemath{\bmtheta}{\boldsymbol{\theta}}
\safemath{\bmphi}{\boldsymbol{\phi}}

\safemath{\bA}{\mathbf{A}}
\safemath{\bB}{\mathbf{B}}
\safemath{\bC}{\mathbf{C}}
\safemath{\bD}{\mathbf{D}}
\safemath{\bE}{\mathbf{E}}
\safemath{\bF}{\mathbf{F}}
\safemath{\bG}{\mathbf{G}}
\safemath{\bH}{\mathbf{H}}
\safemath{\bI}{\mathbf{I}}
\safemath{\bJ}{\mathbf{J}}
\safemath{\bK}{\mathbf{K}}
\safemath{\bL}{\mathbf{L}}
\safemath{\bM}{\mathbf{M}}
\safemath{\bN}{\mathbf{N}}
\safemath{\bO}{\mathbf{O}}
\safemath{\bP}{\mathbf{P}}
\safemath{\bQ}{\mathbf{Q}}
\safemath{\bR}{\mathbf{R}}
\safemath{\bS}{\mathbf{S}}
\safemath{\bT}{\mathbf{T}}
\safemath{\bU}{\mathbf{U}}
\safemath{\bV}{\mathbf{V}}
\safemath{\bW}{\mathbf{W}}
\safemath{\bX}{\mathbf{X}}
\safemath{\bY}{\mathbf{Y}}
\safemath{\bZ}{\mathbf{Z}}

\safemath{\bDelta}{\mathbf{\Delta}}
\safemath{\bLambda}{\mathbf{\Lambda}}
\safemath{\bPhi}{\mathbf{\Phi}}
\safemath{\bSigma}{\mathbf{\Sigma}}
\safemath{\bOmega}{\mathbf{\Omega}}
\safemath{\bTheta}{\mathbf{\Theta}}

\safemath{\bZero}{\mathbf{0}}

\safemath{\veca}{\bma}
\safemath{\vecb}{\bmb}
\safemath{\vecc}{\bmc}
\safemath{\vecd}{\bmd}
\safemath{\vece}{\bme}
\safemath{\vecf}{\bmf}
\safemath{\vecg}{\bmg}
\safemath{\vech}{\bmh}
\safemath{\veci}{\bmi}
\safemath{\vecj}{\bmj}
\safemath{\veck}{\bmk}
\safemath{\vecl}{\bml}
\safemath{\vecm}{\bmm}
\safemath{\vecn}{\bmn}
\safemath{\veco}{\bmo}
\safemath{\vecp}{\bmp}
\safemath{\vecq}{\bmq}
\safemath{\vecr}{\bmr}
\safemath{\vecs}{\bms}
\safemath{\vect}{\bmt}
\safemath{\vecu}{\bmu}
\safemath{\vecv}{\bmv}
\safemath{\vecw}{\bmw}
\safemath{\vecx}{\bmx}
\safemath{\vecy}{\bmy}
\safemath{\vecz}{\bmz}
\safemath{\vecZero}{\bZero}

\safemath{\vecxi}{\bmxi}
\safemath{\veclambda}{\bmlambda}
\safemath{\vecmu}{\bmmu}
\safemath{\vectheta}{\bmtheta}
\safemath{\vecphi}{\bmphi}

\safemath{\matA}{\bA}
\safemath{\matB}{\bB}
\safemath{\matC}{\bC}
\safemath{\matD}{\bD}
\safemath{\matE}{\bE}
\safemath{\matF}{\bF}
\safemath{\matG}{\bG}
\safemath{\matH}{\bH}
\safemath{\matI}{\bI}
\safemath{\matJ}{\bJ}
\safemath{\matK}{\bK}
\safemath{\matL}{\bL}
\safemath{\matM}{\bM}
\safemath{\matN}{\bN}
\safemath{\matO}{\bO}
\safemath{\matP}{\bP}
\safemath{\matQ}{\bQ}
\safemath{\matR}{\bR}
\safemath{\matS}{\bS}
\safemath{\matT}{\bT}
\safemath{\matU}{\bU}
\safemath{\matV}{\bV}
\safemath{\matW}{\bW}
\safemath{\matX}{\bX}
\safemath{\matY}{\bY}
\safemath{\matZ}{\bZ}
\safemath{\matZero}{\bZero}

\safemath{\matDelta}{\bDelta}
\safemath{\matLambda}{\bLambda}
\safemath{\matPhi}{\bPhi}
\safemath{\matSigma}{\bSigma}
\safemath{\matOmega}{\bOmega}
\safemath{\matTheta}{\bTheta}

\newcommand{\hbcomment}[1]{}

\newcommand{\maxo}[1]{\ensuremath{(#1)^{+}}}

\makeatletter
\def\@IEEEinterspaceratioM{0.265}
\def\@IEEEinterspaceMINratioM{0.1651}
\def\@IEEEinterspaceMAXratioM{0.38}

\def\@IEEEinterspaceratioB{0.31}
\def\@IEEEinterspaceMINratioB{0.19}
\def\@IEEEinterspaceMAXratioB{0.38}
\@IEEEtunefonts
\makeatother

\IEEEoverridecommandlockouts

\begin{document}

%AUTHORS
\author{
\begin{tabular}{c}
Cemal Ak\c{c}aba and  Helmut B\"{o}lcskei 
\end{tabular} \\
Communication Technology Laboratory \\
ETH Zurich, Switzerland \\
Email: \{cakcaba {\textbar} boelcskei\}@nari.ee.ethz.ch
}

%TITLE
\title{On the Achievable Diversity-Multiplexing Tradeoff in Interference Channels} 

\maketitle

\begin{abstract}
We analyze two-user single-antenna fading interference channels with perfect receive channel state information (CSI) and no transmit CSI.  For the case of very strong interference, we prove that  decoding interference while treating the intended signal as noise, subtracting the result out, and then decoding the desired signal, a process known as ``stripping'', achieves the diversity-multiplexing tradeoff (DMT) outer bound derived in Akuiyibo and L\'{e}v\^{e}que, \emph{Int. Zurich Seminar on Commun.}, 2008.  The proof is constructive in the sense that it provides corresponding code design criteria for DMT optimality.  For general interference levels, we compute the DMT of a fixed-power-split Han and Kobayashi type superposition coding scheme, provide design criteria for the corresponding superposition codes, and find that this scheme is DMT-optimal for certain multiplexing rates.
\end{abstract}

\section{Introduction}
The interference channel (IC) models the situation where $M$ unrelated transmitters communicate their separate messages to $M$ independent receivers, each of which is assigned to a single transmitter.   Apart from a few special cases \cite{Sato81,Carleial75,Sason04}, the capacity region of the IC remains unknown.  Recently, Etkin \emph{et al.} \cite{Etkin06, Etkin08} showed that in the interference-limited regime\hbcomment{This is terminology used by Etkin \emph{et al.}. They use the term ``interference limited'' to convey the fact that they only consider the cases where signal and interference power is large compared to noise power.}, the capacity region of the IC is achievable to within one bit; later Telatar and Tse \cite{Telatar07} generalized this result to a wider class of ICs. Shang \emph{et al.} derived the noisy-interference sum-rate capacity for Gaussian ICs  in \cite{shangkram08}, while Raja \emph{et al.} \cite{rajavis08} characterized the capacity region of the two-user finite-state compound Gaussian IC to within one bit.  Annapureddy and Veeravalli \cite{veeravalli}  showed that the sum capacity of the two-user Gaussian IC  under weak interference is achieved by treating interference as noise.

In \cite{Akuiyibo08},  Akuiyibo and L\'{e}v\^{e}que derived an outer bound on the diversity-multiplexing tradeoff (DMT) region of fading ICs based on the results of Etkin \emph{et al.} \cite{Etkin08}.   In this paper, we investigate the achievability of this outer bound and we analyze the DMT realized by a stripping decoder and a fixed-power-split Han and Kobayashi (HK)-type superposition coding scheme.  For the sake of simplicity, throughout the paper, we restrict our attention to the two-user case. Furthermore, we assume that the receivers have perfect channel state information (CSI) whereas the transmitters only know the channel statistics.  We would like to point out that the schemes used in \cite{Etkin08} make explicit use of transmit CSI and so does the scheme in \cite{Akuiyibo08}, which immediately implies that the results reported in \cite{Akuiyibo08} serve as an outer bound on the DMT achievable  in the absence of transmit CSI, the case considered here.  The contributions in this paper can be summarized as follows:
\begin{itemize}
\item{For \emph{very strong interference} in the sense of \cite{Etkin08}, we show that a \emph{stripping decoder} which decodes interference while treating the intended signal as noise, subtracts the result out, and then decodes the intended signal is DMT-optimal. We furthermore find that the optimal-DMT can be achieved if each of the two users employs a code that is DMT-optimal on a single-input single-output (SISO) channel.}
\item{For general interference levels, we compute the DMT of a two-message, fixed-power-split HK-type superposition coding scheme and provide design criteria for the corresponding superposition codes. We find that this scheme is DMT-optimal for certain multiplexing rates. \hbcomment{Do we want to be explicit here? Or is it better to say we show it is optimal in `some' cases?}}
\end{itemize}

\subsubsection*{Notation}
The superscripts $^{T}$ and $^{H}$ stand for transpose and conjugate transpose, respectively.
$x_{i}$ represents the $i$th element of the column vector $\mathbf{x}$, and $\lambda_{\min}(\mathbf{X}) $ denotes the smallest eigenvalue of the matrix $\mathbf{X}$.   $\matI_{N}$ is the $N\times N$ identity matrix, and $\mathbf{0}$ denotes the all zeros matrix of appropriate size.  All logarithms are to the base $2$ and $(a)^{+}=\max(a,0)$.  
%and $\Re(x)$ and $\Im(x)$ denote the real and imaginary parts, respectively, of $x \in \mathbb{C}$.
% $\mathcal{A}$  and $\bar{\mathcal{A}}$ denote a set and its complement, respectively. 
$X\sim \mathcal{CN}(0,\sigma^2)$ stands for a circularly symmetric complex Gaussian random variable (RV) with variance $\sigma^{2}$. $f(\rho) \doteq g(\rho)$ denotes exponential equality of the functions $f(\cdot)$ and $g(\cdot)$, i.e., $\lim_{\rho\rightarrow\infty} \log[f(\rho)]/\log \rho  = \lim_{\rho\rightarrow\infty} \log[g(\rho)]/\log \rho.$ 
%\[\lim_{\rho\rightarrow\infty} \frac{\log f(\rho)}{\log \rho}  = \lim_{\rho\rightarrow\infty} \frac{\log g(\rho)}{\log \rho} . \]
The symbols $\dotgeq$, $\dotleq$, $\dot{>}$ and $\dot{<}$ are defined analogously. 
% $\stackrel{d}{=}$ denotes equivalence in distribution. 

\subsubsection*{System model}
We consider a two-user fading IC where two transmitters communicate information to two receivers via a common channel. The fading coefficient between transmitter $i$  $(i=1,2)$ and receiver $j$ $(j=1,2)$ is denoted by $h_{ij}$ and is assumed to be $\mathcal{CN}(0,1)$.  Transmitter $i$ ($ \mathcal{T}_{i}$) chooses an $N$-dimensional codeword $\vecx_{i} \in \mathbb{C}^{N}$, $\|\vecx_{i}\|^{2} \leq N$,  from its codebook, and transmits $\check{\vecx}_{i}= \sqrt{P_{i}}\vecx_{i}$ in accordance with its transmit power constraint $\|\check{\vecx}_{i}\|^{2} \leq NP_{i}$.  In addition, we account for the attenuation of transmit signal $i$ at receiver $j$ ($\mathcal{R}_{j}$) through the real-valued coefficients $\eta_{ij} > 0$.   Defining $\vecy_{i}$ and $\vecz_{i}\sim \mathcal{CN}(\mathbf{0},\matI_{N})$ as the $N$-dimensional received signal vector and noise vector, respectively, at $\mathcal{R}_{i}$, the input-output relation is given by  
 \begin{align}
%\label{intc1}
%	\vecy_{1} &= \eta_{11}h_{11}\check{\vecx}_{1}+\eta_{21}h_{21}\check{\vecx}_{2}+\vecz_{1} \\
\label{intc2}	%\vecy_{2} &= \eta_{12}h_{12}\check{\vecx}_{1}+\eta_{22}h_{22}\check{\vecx}_{2}+\vecz_{2}.
\vecy_{i} &= \eta_{ii}h_{ii}\check{\vecx}_{i}+\eta_{ji}h_{ji}\check{\vecx}_{j}+\vecz_{i} 
\end{align}
for $i,j=1,2$ and $i\neq j$. Setting $\eta_{11}^{2}P_{1} = \eta_{22}^{2}P_{2} = \SNR$ and $\eta_{21}^{2}P_{2} = \eta_{12}^{2}P_{1} = \SNR^{\alpha}$ with $\alpha \in [0,\infty]$ simplifies the exposition and comparison of our results to those in \cite{Etkin08} and \cite{Akuiyibo08}\hbcomment{Your comment here was ``Did \cite{Etkin08} and \cite{Akuiyibo08} also use this specialization of power levels?'' Yes.}\hbcomment{You asked ``Would any of our results change fundamentally if we allowed general power levels at the two transmitters?'' It could be; depending on how one scales interference and signal power levels.}. The resulting equivalent input-output relation is then given by
\begin{align}
\label{intc1s}    \vecy_{i} &= \sqrt{\SNR}h_{ii}\vecx_{i}+\sqrt{\SNR^{\alpha}}h_{ji}\vecx_{j}+ \vecz_{i} 
%\label{intc2s}	\vecy_{2} &= \sqrt{\SNR^{\alpha}}h_{12}\vecx_{1}+\sqrt{\SNR}h_{22}\vecx_{2}+\vecz_{2}.
\end{align} 
for $i,j=1,2$ and $i\neq j$. We assume that both receivers know the signal-to-noise ratio (SNR) value $\SNR$ and the parameter $\alpha$ and $\mathcal{R}_{i}$ $(i=1,2)$ knows $\vech_{i} = [h_{1i} \  h_{2i}]^{T}$ perfectly,  whereas the transmitters only know the channel statistics of all channels\hbcomment{This is required for HK in the optimization step}.  The data rate of $ \mathcal{T}_{i}$ scales with SNR according to $R_{i}=r_{i}\log\SNR$ where the multiplexing rate $r_{i}$ obeys $0 \leq r_{i} \leq 1$. As a result, for $ \mathcal{T}_{i}$ to operate at multiplexing rate $r_{i}$,  we need a sequence of codebooks $\mathcal{C}_{i}(\SNR,r_{i})$, one for each $\SNR$, with $| \mathcal{C}_{i}(\SNR,r_{i})|=2^{NR_{i}}$ codewords $\{\vecx^{1}_{i},\vecx^{2}_{i},\ldots, \vecx_{i}^{2^{NR_{i}}}\}$.  In the following, we will need the multiplexing rate vector $\vecr=[r_{1} \ r_{2}]^{T}$.

\section{Very Strong Interference} 
We call channels with $\alpha \geq 2$ \emph{ very strong interference channels} in the sense of \cite{Etkin08}.  %We shall see that the condition $\alpha \geq 2$ enables each transmitter-receiver pair to communicate as if the interference were not present.  %
Throughout this section, we take $N=1$; we will  see that this results in optimal performance.  In the following, we use the short-hand $x_{i}$ for the first element of the transmit signal vector $\vecx_{i}$, $y_{i}$ for the first element of the receive signal vector $\vecy_{i}$, and $\mathcal{X}_{i}$ for $\mathcal{C}_{i}(\SNR,r_{i})$. 

The error probability corresponding to ML decoding of $ \mathcal{T}_{i}$ at $\mathcal{R}_{i}$ under the assumption that the correctly decoded interference $\mathcal{T}_{j}$ has been removed\hbcomment{This definition is in line with the definition of the outage event} ($i,j=1,2$, $j\neq i$) is denoted by $\prob{E_{ii}|\vech_{i}}$.  We write  $\prob{E_{ij}|\vech_{j}}$ for $i,j=1,2$ and $i\neq j$ for the ML decoding error probability of decoding $ \mathcal{T}_{i}$ at receiver $\mathcal{R}_{j}$ under the assumption that $ \mathcal{T}_{j}$ is treated as noise. The average (w.r.t. the random channel) ML decoding error probability $\Exop_{\vech_{j}}\!\!\left\{\prob{E_{ij}|\vech_{j}}\right\}$ is denoted by $P(E_{ij})$ for $i,j=1,2$. The transmit symbols are assumed equally likely for both transmitters, and hence $\prob{x_{i}} = \frac{1}{|\mathcal{X}_{i}|}$ for $i=1,2$.   The notation $x_{i}^{j}\rightarrow x_{i}^{k}$ represents the event of mistakenly decoding the transmitted codeword $x_{i}^{j} \in \mathcal{X}_{i}$ for the codeword $x_{i}^{k} \in \mathcal{X}_{i}$.  Throughout this section,  as done in \cite{Akuiyibo08}, we use the performance metric $P(E)= \max\{P(E_{11}), P(E_{22})\}$.   The DMT realized by a given  scheme is then characterized by $d(\vecr) = -\lim_{\SNR\rightarrow \infty} \log\left[P(E)\right]/\log\SNR.$
%\begin{align} 
%	d(\vecr) = -\lim_{\SNR\rightarrow \infty}\frac{\log P(E)}{\log\SNR}.
%\end{align}

It is shown in \cite{JournalPrep} that joint decoding of the messages from both transmitters at each receiver achieves the DMT outer bound in \cite{Akuiyibo08} given by $d(\vecr) \leq \min\{(1-r_{1})^{+},(1-r_{2})^{+}\}$. In the following, we show that a stripping decoder also achieves this DMT outer bound. 
%\begin{align}
%\label{Leveq}d(\vecr) \leq \min\{(1-r_{1})^{+},(1-r_{2})^{+}\}.
%  \end{align}

\begin{theorem}
\label{VSint1232}
For the fading IC with I/O relation (\ref{intc1s}), a stripping decoder yields DMT-optimality, i.e., it realizes 
\begin{align}
		P(E) \doteq \SNR^{- \min\{(1-r_{1})^{+},(1-r_{2})^{+}\}}
\end{align}
provided that $\Delta x_{i} = x^{j}_{i}-x_{i}^{k}$ satisfies $|\Delta x_{i}|^{2} \ \dotgeq \ \SNR^{-r_{i}+\epsilon}$ for every pair $x^{j}_{i},x^{k}_{i}$ in each codebook $\mathcal{X}_{i}$, $i=1,2$, and  for some $\epsilon > 0 $. 
\end{theorem}
\begin{proof}
We start by  decoding $ \mathcal{T}_{2}$ at $\mathcal{R}_{1}$ while treating $ \mathcal{T}_{1}$ as noise,  i.e., we have the effective I/O relation
\begin{align}
\label{xzzxe}
	y_{1} & = \sqrt{ \SNR^{\alpha} } h_{21}x_{2} + \tilde{z}
\end{align}
where	$\tilde{z}$ is the effective noise term with variance $1+ \SNR|h_{11}|^{2}$. Recall that $h_{11}$ and $h_{21}$ are known at $\mathcal{R}_{1}$ so that we can condition on $h_{11}$.   We next note that the worst case (in terms of mutual information and hence outage probability) uncorrelated (with the transmit signal) additive noise under a variance constraint is Gaussian \cite[Theorem 1]{Hassibi03}.  In the following, we use the corresponding worst-case outage probability to exponentially upper-bound $P(E_{21})$, i.e., we set $\tilde{z}\sim \mathcal{CN}(0, 1+ \SNR|h_{11}|^{2})$.   We start by normalizing the received signal according to
\begin{align}
	\frac{y_{1}}{\sqrt{1+\SNR|h_{11}|^2}} & = \sqrt{\frac{\SNR^{\alpha}}{1+\SNR|h_{11}|^2}}h_{21}x_{2}+ z
\end{align}	
where $ z \sim \mathcal{CN}(0,1)$.  We can now upper-bound $\prob{E_{21}|\vech_{1}}$ as
\begin{align}
&	\prob{E_{21}|\vech_{1}} = \sum_{x_{2} \in \mathcal{X}_{2}} \prob{x_{2}}\prob{E_{21}| \vech_{1}, x_{2} } \\
%\label{equipp}							    & \ \ \ = \frac{1}{|\mathcal{X}_{2}|} \sum_{i=1}^{|\mathcal{X}_{2}|} \prob{\bigcup\limits_{\substack{ j =1    \\ {j  \neq i}}}^{|\mathcal{X}_{2}|}  \! x^{i}_{2} \rightarrow x^{j}_{2}\left.\right| \vech_{1}} \\
\label{unionbound}							    & \ \ \ \leq |\mathcal{X}_{2}| \prob{x^{\tilde{i}}_{2} \rightarrow x^{\tilde{j}}_{2}\left.\right| \vech_{1}} \\
							 \label{cortez}   &\ \ \ \leq |\mathcal{X}_{2}| {\rm Q} \left(\sqrt{\frac{\SNR^{\alpha}|h_{21}|^{2}|\Delta x_{2}|^{2}}{2(1+\SNR|h_{11}|^{2})}}\right)
\end{align}
where $\left\{x_{2}^{\tilde{i}}, x_{2}^{\tilde{j}}\right\}$ denotes the (or ``a'' in the case of multiple pairs with the same distance) pair of symbols with minimum Euclidean distance among all possible pairs of different symbols.  We next define the outage event $\mathcal{O}_{ii}$ associated with decoding $ \mathcal{T}_{i}$ at $\mathcal{R}_{i}$ ($i=1,2$) in the absence of  interference and its complementary event $\bar{\mathcal{O}}_{ii}$  as follows
\begin{align}
\label{fax}	\mathcal{O}_{ii} &= \left\{h_{ii}: \log\left(1+\SNR|h_{ii}|^{2}\right) < R_{i} \right\} \\
\label{faxoz}	\bar{\mathcal{O}}_{ii} &= \left\{h_{ii}: \log\left(1+\SNR|h_{ii}|^{2}\right) \geq R_{i} \right\}.
\end{align}
We note that the definitions \eqref{fax} and \eqref{faxoz} are in line with the definitions of $P(E_{ii})$ for $i=1,2$.  Similarly, we define the event $\mathcal{O}_{ij}$ associated with decoding $ \mathcal{T}_{i}$ at $\mathcal{R}_{j}$ while treating $ \mathcal{T}_{j}$ as noise ($i,j=1,2$ and $i\neq j$) and its complementary event $\bar{\mathcal{O}}_{ij}$  as follows
\begin{align}
	\mathcal{O}_{ij} &= \left\{\vech_{j}: \log\left(1+\frac{\SNR^{\alpha}|h_{ij}|^{2}}{1+\SNR|h_{jj}|^2}\right) < R_{i} \right\} \notag \\
	\bar{\mathcal{O}}_{ij} &=  \left\{\vech_{j}: \log\left(1+\frac{\SNR^{\alpha}|h_{ij}|^{2}}{1+\SNR|h_{jj}|^2}\right) \geq R_{i} \right\} \notag. 
\end{align}
Next, we upper-bound $P(E_{21})$ according to
\begin{align}
&P(E_{21}) = \Exop_{\vech_{1}} \! \lefto\{\prob{E_{21}|\vech_{1}}\right\}  =  \notag \\[0.1cm]%=  \Exop_{\vech_{1}}\lefto\{\prob{E_{21},\mathcal{O}_{21}} +\prob{E_{21},\bar{\mathcal{O}}_{21}} \right\}  \\
& \label{molt1}  \Exop_{\vech_{1}}\! \lefto\{\prob{\mathcal{O}_{21}}\prob{E_{21}|\vech_{1}, \mathcal{O}_{21}} \! + \!  \prob{\bar{\mathcal{O}}_{21}}\prob{E_{21}|\vech_{1},\bar{\mathcal{O}}_{21}} \! \right\}   \\[0.1cm]
 &\label{molt2}  \ \ \ \ \ \leq  \prob{\mathcal{O}_{21}}+ \Exop_{\vech_{1}}\! \lefto\{\prob{E_{21}|\vech_{1},\bar{\mathcal{O}}_{21}} \right\}  \\[0.1cm]
 &\label{molt3} \ \ \ \  \ \leq \prob{\mathcal{O}_{21}} + \SNR^{r_{2}} {\rm Q} \left(\sqrt{\frac{\SNR^{r_{2}}|\Delta x_{2}|^{2}}{2}}\right)
\end{align} where (\ref{molt1}) follows from Bayes's rule and (\ref{molt2}) is obtained by upper-bounding $\prob{E_{21}|\vech_{1}, \mathcal{O}_{21}}$ and $\prob{\bar{\mathcal{O}}_{21}}$ by $1$.  Finally, (\ref{molt3}) follows by using the fact that $\bar{\mathcal{O}}_{21}$ implies $\frac{\SNR^{\alpha}|h_{21}|^{2}}{1+\SNR|h_{11}|^2} \geq 2^{R_{2}}-1$, and invoking $R_{2}=r_{2}\log\SNR$, $|\mathcal{X}_{2}| = \SNR^{r_{2}}$, and $\SNR \gg 1$ in (\ref{cortez}).  It can be shown that $\prob{\mathcal{O}_{21}} \doteq \SNR^{-(\alpha-1-r_{2})^{+}}$ for $\alpha \geq  2$ \cite{Akuiyibo08}.  Further, since $|\Delta x_{2}|^{2} \ \dotgeq \ \SNR^{-r_{2}+\epsilon}$, for  $\epsilon > 0$, by assumption,  we can further simplify the above as the second term in (\ref{molt3}) decays exponentially in SNR whereas the first term decays polynomially, i.e., %\begin{align}
$\Exop_{\vech_{1}}\! \left\{\prob{E_{21}|\vech_{1}}\right\}  \ \dotleq \  \prob{\mathcal{O}_{21}} \doteq \SNR^{-(\alpha-1-r_{2})^{+}}.$
%\end{align}
We proceed to analyze decoding of $ \mathcal{T}_{1}$ at $\mathcal{R}_{1}$ and start by defining $\bar{x}_{2}$ as the result of decoding $ \mathcal{T}_{2}$ at $\mathcal{R}_{1}$.  Note that we do not need to assume that $ \mathcal{T}_{2}$ was decoded correctly at $\mathcal{R}_{1}$. We begin by upper-bounding $\prob{E_{11}|\vech_{1}}$ given $\bar{x}_{2}$: 
\begin{align}
&\prob{E_{11}|\vech_{1},\bar{x}_{2}}   \notag= \sum_{x_{1} \in \mathcal{X}_{1}} \sum_{x_{2}\in \mathcal{X}_{2}} \prob{x_{1}} \prob{x_{2}} \prob{E_{1}| \vech_{1},x_{1}, x_{2}, \bar{x}_{2}} \\ 
%\label{xxza2}& = \sum_{x_{1} \in \mathcal{X}_{1}} \sum_{x_{2}\in \mathcal{X}_{2}} \prob{x_{1}} \prob{x_{2}} \prob{E_{1}| \vech_{1},x_{1}, x_{2}, \bar{x}_{2}} \\ 
%\label{xxza3}& = \frac{1}{|\mathcal{X}_{1}||\mathcal{X}_{2}|} \sum_{i=1}^{|\mathcal{X}_{1}|}   \sum_{k=1}^{|\mathcal{X}_{2}|}  \prob{\!  \bigcup\limits_{\substack{j =1     \\ {j\neq i}}}^{|\mathcal{X}_{1}|}  x^{i}_{1} \! \rightarrow x^{j}_{1}\left.\right|\vech_{1},x_{2}^{k},\bar{x}_{2}}  \\
\label{xxza4}& \leq \frac{|\mathcal{X}_{1}|}{|\mathcal{X}_{2}|}\sum_{k=1}^{|\mathcal{X}_{2}|}\prob{x^{\tilde{i}}_{1} \rightarrow x^{\tilde{j}}_{1}\!\left.\right|\vech_{1},x_{2}^{k},\bar{x}_{2}}
\end{align} where $\left\{x_{1}^{\tilde{i}}, x_{1}^{\tilde{j}}\right\}$ denotes the (or ``a'' in the case of multiple pairs with the same distance) pair of symbols with minimum Euclidean distance among all possible pairs of different symbols. Next, we further upper-bound $\prob{E_{11}|\vech_{1},\bar{x}_{2}}$ by considering two events; namely, when $\mathcal{R}_{1}$ decodes $ \mathcal{T}_{2}$ correctly and when it does not:
\begin{align}
& \prob{E_{11}|\vech_{1},\bar{x}_{2}}  \leq  \notag \\ 
& \frac{|\mathcal{X}_{1}|}{|\mathcal{X}_{2}|} \sum_{k=1}^{|\mathcal{X}_{2}|} \left(\prob{\bar{x}_{2}\!=x_{2}^{k}| \vech_{1}, x_{2}^{k}}\!\prob{x^{\tilde{i}}_{1} \!  \rightarrow x^{\tilde{j}}_{1}  \!\left.\right| \! \vech_{1},x_{2}^{k}, \bar{x}_{2} ,\bar{x}_{2} \! = \! x_{2}^{k}} \! \right.  \notag \\
&\left. + \prob{\bar{x}_{2}\! \neq x_{2}^{k}|\vech_{1}, x_{2}^{k}}\!\prob{x^{\tilde{i}}_{1} \!  \rightarrow x^{\tilde{j}}_{1}\!  \left. \right| \! \vech_{1},x_{2}^{k}, \bar{x}_{2}, \bar{x}_{2}\! \neq x_{2}^{k}} \!  \right) \! , \label{junkf}
\end{align}
where $\prob{x^{\tilde{i}}_{1} \!  \rightarrow x^{\tilde{j}}_{1}\! \left.\right| \! \vech_{1},x_{2}^{k}, \bar{x}_{2} ,\bar{x}_{2} \! = \! x_{2}^{k}}$ is the probability of mistakenly decoding $x_{1}^{\tilde{i}}$ for $x_{1}^{\tilde{j}}$ given that $ \mathcal{T}_{2}$ transmitted $x_{2}^{k}$ and $\mathcal{R}_{1}$ decoded $ \mathcal{T}_{2}$ correctly, i.e.,  $\bar{x}_{2} =x_{2}^{k}$.  The quantity $\prob{\bar{x}_{2} = x_{2}^{k}|\vech_{1}, x_{2}^{k}}$ is the probability of decoding $ \mathcal{T}_{2}$ correctly given that $x_{2}^{k}$ was transmitted. By upper-bounding $\prob{\bar{x}_{2}=x_{2}^{k}|\vech_{1},x_{2}^{k}}$ and $\prob{x^{\tilde{i}}_{1} \! \rightarrow x^{\tilde{j}}_{1}\left.\right| \! \vech_{1},x_{2}^{k}, \bar{x}_{2}, \bar{x}_{2} \! \neq x_{2}^{k}}$ in (\ref{junkf}) by $1$, we arrive at 
\begin{align}
 \prob{E_{11}|\vech_{1},\bar{x}_{2}} &\leq \frac{|\mathcal{X}_{1}|}{|\mathcal{X}_{2}|} \sum_{k=1}^{|\mathcal{X}_{2}|} \prob{x^{\tilde{i}}_{1} \rightarrow x^{\tilde{j}}_{1} \! \left.\right|\vech_{1},x_{2}^{k},\bar{x}_{2},\bar{x}_{2}=x_{2}^{k}} + \notag \\   
 &\label{georg}  \ \ \ \ \ \ \ \  \ \  \frac{|\mathcal{X}_{1}|}{|\mathcal{X}_{2}|}\sum_{k=1}^{|\mathcal{X}_{2}|} \prob{\bar{x}_{2}\neq x_{2}^{k}|\vech_{1}, x_{2}^{k}}. 
\end{align}
Next, noting that\hbcomment{$\frac{1}{|\mathcal{X}_{2}|}\sum\limits_{k=1}^{|\mathcal{X}_{2}|}\prob{\bar{x}_{2}\neq x_{2}^{k}|\vech_{1}, x_{2}^{k}}$ is the average error probability, and is minimized by the ML receiver that treats $x_{1}$ as discrete noise} $\frac{1}{|\mathcal{X}_{2}|}\sum\limits_{k=1}^{|\mathcal{X}_{2}|}\prob{\bar{x}_{2}\neq x_{2}^{k}|\vech_{1}, x_{2}^{k}} \leq \prob{E_{21}|\vech_{1}}$ and invoking the corresponding upper bound (\ref{cortez}) in (\ref{georg}), we get
 \begin{align}
 \prob{E_{11}|\vech_{1},\bar{x}_{2}} & \leq |\mathcal{X}_{1}|{\rm Q}\left(\sqrt{\frac{\SNR|h_{11}|^{2}|\Delta  x_{1}|^{2}}{2}}\right)+ \notag \\ & \ \ \ \ 
\label{georg2}|\mathcal{X}_{1}||\mathcal{X}_{2}|{\rm Q}\left(\sqrt{\frac{\SNR^{\alpha}|h_{21}|^{2}|\Delta  x_{2}|^{2}}{2(1+\SNR|h_{11}|^{2})}}\right).
\end{align} The first term on the RHS of (\ref{georg2}) follows from the first term on the RHS of (\ref{georg}), since given $\bar{x}_{2} =x_{2}^{k}$, the interference is subtracted out perfectly, leaving an effective SISO channel without interference.  We are now in a position to upper-bound $P(E_{11})$:
\begin{align}
\label{consxzzz}	& P(E_{11}) = \Exop_{\vech_{1}}\!\lefto\{\prob{E_{11}|\vech_{1}}\right\} \leq   \Exop_{\vech_{1}}\!\lefto\{\prob{E_{11}|\vech_{1},\bar{x}_{2}}\right\} \\[0.12cm] 
	& \leq \Exop_{\vech_{1}}\lefto\{ |\mathcal{X}_{1}|{\rm Q}\left(\sqrt{\frac{\SNR|h_{11}|^{2}|\Delta  x_{1}|^{2}}{2}}\right)\right\}+\notag \\[0.12cm] 
\label{xxzassr}	& \Exop_{\vech_{1}}\lefto\{|\mathcal{X}_{1}||\mathcal{X}_{2}|{\rm Q}\left(\sqrt{\frac{\SNR^{\alpha}|h_{21}|^{2}|\Delta  x_{2}|^{2}}{2(1+\SNR|h_{11}|^{2})}}\right) \right\}.
\end{align}
Here, (\ref{consxzzz}) follows since the error probability incurred by using the stripping decoder constitutes a natural upper bound on $\Exop_{\vech_{1}}\!\lefto\{\prob{E_{11}|\vech_{1}}\right\} $. We upper-bound (\ref{xxzassr}) by splitting each of the two terms into outage and no outage sets using Bayes's rule to arrive at
\begin{align}
	&  P(E_{11}) = \Exop_{\vech_{1}}\lefto\{\prob{E_{11}|\vech_{1}}\right\} \leq  \notag \\[0.08cm] 
 	& \prob{\mathcal{O}_{11}} + \SNR^{r_{1}}{ \rm Q}\lefto(\sqrt{\frac{\SNR^{r_1}|\Delta x_{1}|^{2}}{2}}\right) + \prob{\mathcal{O}_{21}} + \notag 
	\\ 	& \ \ \ \ \ \ \ \ \ \ \ \ \ \ \ \ \ \ \ \ \ \ \ \ \ \ \ \ \ \ \ \ \   
\label{barbu}	\SNR^{r_{1}+r_{2}}{ \rm Q}\lefto(\sqrt{\frac{\SNR^{r_2}|\Delta x_{2}|^{2}}{2}}\right). 
 \end{align}
The second and fourth terms on the RHS of (\ref{barbu}) follow from (\ref{xxzassr}) since $\bar{\mathcal{O}}_{11}$ and $\bar{\mathcal{O}}_{21}$ imply $\SNR |h_{11}|^{2} \geq 2^{R_{1}}-1$ and $\frac{\SNR^{\alpha}|h_{21}|^{2}}{1+\SNR|h_{11}|^2} \geq 2^{R_{2}}-1$, respectively, and since $R_{i}=r_{i}\log\SNR$, $|\mathcal{X}_{i}| = \SNR^{r_{i}}$ for $i=1,2 $ and $\SNR \gg 1$.  Given that the minimum Euclidean distances in each codebook, $|\Delta  x_{1}|^{2}$ and $|\Delta  x_{2}|^{2}$,  obey $|\Delta  x_{1}|^{2} \ \dotgeq \ \SNR^{-r_{1}+\epsilon}$ and $|\Delta  x_{2}|^{2} \ \dotgeq \ \SNR^{-r_{2}+\epsilon}$, for some $\epsilon > 0$, by assumption, we get
\begin{align}
	P(E_{11}) &= \Exop_{\vech_{1}}\!\left\{\prob{E_{11}|\vech_{1}}\right\}  \ \dotleq  \ \ \prob{\mathcal{O}_{11}} +  \prob{\mathcal{O}_{21}} \\ 
											                  & \ \doteq  \  \SNR^{-\maxo{1-r_{1}}} + \SNR^{-\maxo{\alpha-1-r_{2}}} \\
\label{xxxaszadv1}											                  &  \ \doteq  \  \SNR^{-\min\{\maxo{1-r_{1}},\maxo{\alpha-1-r_{2}}\}}.
\end{align}	
Similar derivations for decoding at $\mathcal{R}_{2}$ lead to  
%\begin{align} \label{xxxaszadv2}			
$P(E_{22})            \ \dotleq  \  \SNR^{-\min\{\maxo{1-r_{2}},\maxo{\alpha-1-r_{1}}\}}.$
%\end{align}
We note that the error probability of decoding $ \mathcal{T}_{i}$ at $\mathcal{R}_{i}$ is exponentially lower-bounded by $\prob{\mathcal{O}_{ii}}$ for $i=1,2$ \cite{zheng_tradeoff}.  Hence, $P(E_{ii})$ is  sandwiched according to
\begin{align}
\label{sand1} &\!\!\! \SNR^{-\maxo{1-r_{i}}}  \dotleq \  P(E_{ii})  \ \dotleq   \  \SNR^{-\min\{\maxo{1-r_{i}},\maxo{\alpha-1-r_{j}}\}} 
\end{align}
for $i,j=1,2$ and $i\neq j$.  The proof is concluded by first upper-bounding $P(E) =\max \lefto\{P(E_{11}), P(E_{22})\right\}$ as
\begin{align}
P(E)  \ \dotleq  \ & \max  \lefto\{\SNR^{-\min\{\maxo{1-r_{1}},\maxo{\alpha-1-r_{2}}\}}, \right. \notag  \\ 
 & \ \ \ \left.   \SNR^{-\min\{\maxo{1-r_{2}},\maxo{\alpha-1-r_{1}}\}}\right\}  \notag \\
\label{sensib1} \doteq & \ \SNR^{-\min\lefto\{\maxo{1-r_{1}},\maxo{1-r_{2}}\right\}}
\end{align}
where  (\ref{sensib1}) is a consequence of the assumption $\alpha \geq 2$.   Secondly, $P(E)$ can be lower-bounded using the outage bounds on the individual error probabilities $P(E_{11})$ and $P(E_{22})$:
\begin{align}
\label{sensib2}	\SNR^{-\min\lefto\{\maxo{1-r_{1}},\maxo{1-r_{2}}\right\}} \ &\dotleq  \ P(E).	
\end{align}
Since the $\SNR$ exponents in the upper bound (\ref{sensib1}) and the lower bound (\ref{sensib2}) match, we can conclude that  % \begin{align}
$P(E) \doteq \SNR^{-\min\lefto\{\maxo{1-r_{1}},\maxo{1-r_{2}}\right\}}$
 %\end{align}
which establishes the desired result. \end{proof}
\begin{remark} 
We can immediately conclude from Theorem 1 that using a sequence of codebooks that is DMT-optimal for the SISO channel for both users results in DMT-optimality for the IC under very strong interference. 
\end{remark}
\begin{remark} 
If $R_{1}=R_{2}=r\log\SNR$ and we use sequences of codebooks $\mathcal{C}(\SNR,r)$ satisfying the conditions of Theorem 1 for both users,  then we have $P(E_{11}) \doteq P(E_{22}) \doteq  \SNR^{-(1-r)^{+}}$ %\begin{align}
%P(E_{11}) \doteq P(E_{22}) \doteq  \SNR^{-(1-r)^{+}}
%\end{align}
as a simple consequence of (\ref{sand1}). This means that in the special case, where each $ \mathcal{T}_{i}$ transmits at the same multiplexing rate $r$, we have the stronger result that the single user DMT, i.e., the DMT that is achievable for a SISO channel in the absence of any interferers,  is achievable for both users.  In effect, under very strong interference and when the two users operate at the same multiplexing rate, the interference channel effectively gets \emph{decoupled}. For $r_{1}\neq r_{2}$, we can, in general, not arrive at the same conclusion as the SNR exponents in (\ref{sand1}) do not necessarily match. Joint decoding at both receivers is, however, shown in  \cite{JournalPrep} to decouple the very strong interference (fading) channel for $\alpha \geq 2$ for all values of $r_{i}$, $i=1,2$, i.e., for $0 \leq r_{i} \leq 1$, $i=1,2$. 
\end{remark}

\section{General Interference Channels and Han and Kobayashi Schemes} 
The HK rate region \cite{hanko} remains the best known achievable rate region for the Gaussian IC \cite{Sason04, kramer06}.  The original HK strategy lets each transmitter split its message into two messages, and allows each receiver to decode part of the interfering signal. %The original HK scheme uses five auxiliary RVs $Q, U_{1}, U_{2}, W_{1},$ and $W_{2},$ all defined on arbitrary finite sets.  The auxiliary RV $U_{i}$ carries the private message of $ \mathcal{T}_{i}$, whereas the auxiliary RV $W_{i}$ carries the public message of $ \mathcal{T}_{i}$ destined for both receivers.  The RV $Q$ is for time-sharing.    The general HK region is usually prohibitively complex to describe \cite{Chong08}. 

In the following, we analyze the DMT of a superposition HK scheme where $ \mathcal{T}_{i}$ transmits the $N$-dimensional\hbcomment{Is it integral that we comment on $N\geq 2$ here?} ($N \geq 2$) vector $\vecx_{i} = \vecu_{i}+\vecw_{i}$ with $\vecu_{i}$ and $\vecw_{i}$ representing the private and the public message, respectively.   All assumptions of Section I remain valid, and we allow all levels of interference, i.e., $\alpha \geq 0$.  The power constraints for $\vecu_{i}$ and $\vecw_{i}$ are  
\begin{align}
\| \vecu_{i} \| \leq \sqrt{\frac{N}{\SNR^{1-p_{i}}}},  \ \  
\| \vecw_{i} \| \leq \sqrt{N}\lefto(1-\sqrt{\frac{1}{\SNR^{1-p_{i}}}}\right)  \notag
\end{align}
so that $\|\vecx_{i}\| \leq \|\vecu_{i}\| + \|\vecw_{i}\| = \sqrt{N}$.  Here, $0 \leq p_{i} < 1$ accounts for  the exponential order of the power allocated to the private message. The power split is assumed fixed and is independent of the channel realizations.  When both the private and the public message are allocated maximum power, we have $\frac{\|\vecw_{i}\|^{2}}{\|\vecu_{i}\|^{2}} \doteq \SNR^{1-p_{i}}$.  %We emphasize that any $p_{i} < 1$  constitutes a valid power split.  Schemes with $p_{i} < 0 $  yield zero diversity order, and, hence, do not contribute to the DMT region \cite{JournalPrep}. 
We emphasize that any $p_{i} < 1$  constitutes a valid power split.  We explain in \cite{JournalPrep} why we can omit all the cases with $p_{i}<0$ except for  $p_{i} =-\infty$. The special case $p_{i}=-\infty$ corresponds to using public messages only and is, therefore,  similar to a multiuser setup. This case  is treated separately in the following and is referred to by the subscript $_{MU}$. 

We assume that $ \mathcal{T}_{i}$ transmits at rate $R_{i}=r_{i}\log \SNR$  where  the rates for the private and the public messages, respectively,   are  $S_{i}=s_{i} \log \SNR$ and $T_{i} = t_{i} \log \SNR$ with $r_{i}= s_{i}+t_{i}$, $s_{i}, t_{i}\geq 0$, and $0 \leq r_{i}\leq 1$.  The codebooks corresponding to the private and the public message parts are denoted as  $\mathcal{C}^{\vecu_{i}}(\SNR,s_{i})$ and $\mathcal{C}^{\vecw_{i}}(\SNR,t_{i})$, respectively, and satisfy $|\mathcal{C}^{\vecu_{i}}(\SNR,s_{i})|=\SNR^{Ns_{i}}$ and $|\mathcal{C}^{\vecw_{i}}(\SNR,t_{i})|=\SNR^{Nt_{i}}$. Clearly, $\mathcal{C}^{\vecx_{i}}(\SNR,r_{i}) = \mathcal{C}^{\vecu_{i}}(\SNR,s_{i})\times \mathcal{C}^{\vecw_{i}}(\SNR,t_{i}) $ with $|\mathcal{C}^{\vecx_{i}}(\SNR,r_{i})| = \SNR^{r_{i}}$.   In the following, we will need the private message multiplexing rate vector  $\vecs=[s_{1} \ s_{2}]^{T}$ and the SNR exponent vector $\vecp= [p_{1} \ p_{2}]^{T}$ of the private messages.  As before, our performance metric is  $P(E) = \max\{P(E_{11}), P(E_{22}) \}$.   
\begin{theorem}
The maximum DMT achievable by a fixed-power-split HK scheme is given by 
\begin{align} 
d(\vecr) = \max\lefto\{d_{HK}(\vecr), d_{MU}(\vecr)\right\} 
\end{align}
where $d_{MU}(\vecr) =   \min\limits_{i=1,2,3}\left\{d_{MU}^{i}(\vecr)\right\}$  with 
 \begin{align}
	&d_{MU}^{i}(\vecr)= (1-r_{i})^{+} \ \ \ \ \text{for} \ \ \  i=1,2 \\
	& d_{MU}^{3}(\vecr) = \lefto(1-r_{1}-r_{2}\right)^{+}+ \lefto(\alpha-r_{1}-r_{2}\right)^{+} \notag
\end{align}
and 
	\begin{align}
\label{dhk}d_{HK}(\vecr) = \max_{\vecs,\vecp} d(\vecr,\vecs, \vecp) 
\end{align} where the optimization is carried out subject to the constraints $ s_{i}+t_{i} = r_{i}$ with  $s_{i},t_{i} \geq 0$, and 
	$0 \leq p_{i}  < 1$, all for $i =1,2$, and
\begin{align} d(\vecr,\vecs, \vecp)  &= \min_{\substack{ k=1,2 \\ l=1,2,\ldots,6}}\left\{d_{kl}(\vecr,\vecs, \vecp)\right\}   \notag \\
d_{i1}(\vecr,\vecs,\vecp) &= 	\begin{cases} 
	(p_{i}-s_{i})^{+}, &   \text{if } \   p_{j} <  1-\alpha \\
	(1-\alpha-p_{j}+p_{i}-s_{i})^{+}, &  \text{if } \   p_{j} \geq  1 - \alpha
	\end{cases} \notag \\
d_{i2}(\vecr,\vecs,\vecp)  &= \begin{cases} 
	(1-r_{i}+s_{i})^{+}, & \  \text{if } \   p_{j}  <  1-\alpha   \\
	(2-\alpha-p_{j}-r_{i}+s_{i})^{+}, & \ \text{if } \   p_{j} \geq  1-\alpha  
	\end{cases} \notag \\
d_{i3}(\vecr,\vecs,\vecp) &= 	\begin{cases} 
	(1-r_{i})^{+}, & \  \text{if } \   p_{j}  <  1-\alpha \notag \\
	(2-\alpha-p_{j}-r_{i})^{+}, & \ \text{if } \   p_{j} \geq  1-\alpha \notag
	\end{cases} \\
d_{i4}(\vecr,\vecs,\vecp) &= 	\begin{cases} 
	(p_{i}-s_{i}-r_{j}+s_{j})^{+} \! + \! (\alpha-s_{i}-r_{j}+s_{j})^{+}, \\  \ \ \ \ \ \ \ \ \ \ \ \ \  \text{if } \   p_{j} < 1-s_{i}-r_{j}+s_{j} \notag \\
	(p_{i}-s_{i}-r_{j}+s_{j})^{+}, \\ \ \ \ \ \ \text{if } \   p_{j} \geq  1-s_{i}-r_{j}+s_{j}  \ \text{and} \ p_{j} < 1-\alpha \notag \\
	(1-\alpha-p_{j}+p_{i}-s_{i}-r_{j}+s_{j})^{+}, \\ \ \ \ \ \ \text{if } \   p_{j} \geq  1-s_{i}-r_{j}+s_{j} \ \text{and} \ p_{j} \geq 1-\alpha \notag 
	\end{cases} \\
d_{i5}(\vecr,\vecs,\vecp) &=  \begin{cases} 
	\lefto(\! 1- \! \! \sum\limits_{k=1}^{2}r_{k}+\! \sum\limits_{l=1}^{2} s_{l} \! \right)^{\! \! +}\! \! +\! \lefto(\! \alpha -\! \sum\limits_{k=1}^{2}r_{k}+\! \sum\limits_{l=1}^{2} s_{l} \! \right)^{\! \! +}\! ,  \\  \ \  \text{if } \   p_{j} <  1-\sum\limits_{k=1}^{2}r_{k}+\sum\limits_{l=1}^{2} s_{l}, \\
	\lefto(\! 1-\sum\limits_{k=1}^{2}r_{k}+\sum\limits_{l=1}^{2} s_{l} \! \right)^{\! \! +},  \\  \ \  \text{if } \   p_{j} \geq  1-\sum\limits_{k=1}^{2}r_{k}+\sum\limits_{l=1}^{2} s_{l} \ \text{and} \ p_{j} < 1-\alpha \notag \\
	\lefto(2-\alpha-p_{j}-\sum\limits_{k=1}^{2}r_{k}+\sum\limits_{l=1}^{2} s_{l}\right)^{\! \! +},  \\  \ \  \text{if } \   p_{j} \geq  1-\sum\limits_{k=1}^{2}r_{k}+\sum\limits_{l=1}^{2} s_{l} \ \text{and} \ p_{j} \geq 1-\alpha
	\end{cases} \\	
d_{i6}(\vecr,\vecs,\vecp) &=	\begin{cases} 
	(1-r_{i}-r_{j}+s_{j})^{+}+(\alpha-r_{i}-r_{j}+s_{j})^{+},  \\ \ \ \ \ \ \ \ \ \ \ \ \ \ \ \    \text{if } \   p_{j} <  1-r_{i}-r_{j}+s_{j} \notag \\
	(1-r_{i}-r_{j}+s_{j})^{+},  \\ \ \ \ \ \ \text{if } \   p_{j} \geq  1-r_{i}-r_{j}+s_{j} \ \text{and} \ p_{j} < 1-\alpha  \notag \\
	(2-\alpha-p_{j}-r_{i}-r_{j}+s_{j})^{+},  \\  \ \ \ \ \ \text{if } \   p_{j} \geq  1-r_{i}-r_{j}+s_{j} \ \text{and} \ p_{j} \geq 1-\alpha\notag
	\end{cases}
\end{align} with $i,j=1,2$ and $i\neq j$.
\end{theorem} We shall next provide code design criteria for achieving the DMT in Theorem 2. 
\begin{theorem}	
For a given rate tuple $\vecr$, either $d_{HK}(\vecr)$ or $d_{MU}(\vecr)$ dominates in Theorem 2.

\begin{enumerate}
\item{  If $d_{HK}(\vecr) \leq d_{MU}(\vecr)$,  the DMT in Theorem 2 is achieved as follows.  Denote $j^{*}= \arg \min_{i=1,2,3} d_{MU}^{i}(\vecr)$.  Let $\Gamma_{i}(\vecr) =[\gamma_{i}^{1}(\vecr) \ \gamma_{i}^{2}(\vecr)]^{T}$ be the functions\hbcomment{We note that functions $\Gamma_{i}(\vecr)$ might not be unique.} such that $d_{MU}^{j^{*}}(\vecr) = d_{MU}^{i}(\Gamma_{i}(\vecr))$  for $i=1,2,3$.  Then, the DMT in Theorem 2 is achieved by employing a sequence (in SNR) of codebooks satisfying
\begin{align}
	\| \Delta \vecx_{i}\|^{2} \ & \dotgeq \ \SNR^{-\gamma_{i}^{i}(\vecr)+\epsilon}, \label{xxczzxvz}\\
	\lambda_{\min}\lefto(\Delta\matX (\Delta\matX)^{H}\right) \ & \dotgeq \ \SNR^{-\gamma_{3}^{1}(\vecr)-\gamma_{3}^{2}(\vecr)+\epsilon} \label{xxczzxvz2}
\end{align}
for some\footnote{We note that all $\epsilon$'s in \eqref{xxczzxvz}-\eqref{xxczzxvz2} and  equation block \eqref{faflafl} can be different.} $\epsilon >  0$, where  $\Delta \vecx_{i} = \hat{\vecx}_{i} - \tilde{\vecx}_{i}$, $i = 1,2$,  with $\hat{\vecx}_{i}, \tilde{\vecx}_{i} \in \mathcal{C}^{\vecx_{i}}(\SNR,r_{i})$, and $\Delta\matX = [\Delta\vecx_{1} \ \Delta\vecx_{2}]$.}

\item{ If $d_{HK}(\vecr) > d_{MU}(\vecr)$,   then define the codeword difference vectors $\Delta\vecu_{i}=\sqrt{\SNR^{1-p_{i}}}(\tilde{\vecu}_{i}-\hat{\vecu}_{i})$, $\Delta\vecw_{i}=\tilde{\vecw}_{i}-\hat{\vecw}_{i}$, and $\Delta\vecx_{i}=\tilde{\vecx}_{i}-\hat{\vecx}_{i}$ with $\tilde{\vecu}_{i},\hat{\vecu}_{i} \in \mathcal{C}^{\vecu_{i}}(\SNR, s_{i})$, $\tilde{\vecw}_{i},\hat{\vecw}_{i} \in \mathcal{C}^{\vecw_{i}}(\SNR,t_{i})$ and $\tilde{\vecx}_{i},\hat{\vecx}_{i} \in \mathcal{C}^{\vecx_{i}}(\SNR, r_{i})$,  for $i=1,2$. Further, define  $\Delta\matA_{ij} = [ \Delta\vecu_{i}  \ \Delta \vecw_{j}]$, $\Delta\matB_{ij} = [ \Delta\vecw_{i}  \ \Delta \vecw_{j}]$,  and $\Delta\matC_{ij} = [ \Delta\vecx_{i}  \ \Delta \vecw_{j}]$ for $i,j=1,2$ and $i\neq j$.  Denote the optimizing values of $\vecs$, $\vect$, and $\vecp$  obtained by solving (\ref{dhk}) as $\vecs^{*}, \vect^{*}$, and $\vecp^{*}$, respectively. %Note that $\vect^{*}$ follows from $\vecs^{*}$, since $\vecr$ is fixed and $\vecr = \vecs^{*}+\vect^{*}$. 
We let 
\begin{align} [k^{*} \ l^{*}] = \argmin_{\substack{ k=1,2 \\ l=1,2,3,4,5,6}}\left(d_{kl}(\vecr,\vecs, \vecp)\right).\end{align} 
Further, let the functions\hbcomment{We note that functions $\Upsilon_{nm}(\vecr)$ and $\Psi_{nm}(\vecs^{*})$ might not be unique.} $\Upsilon_{nm}(\vecr)=[\upsilon_{nm}^{1}(\vecr) \ \upsilon_{nm}^{2}(\vecr)]^{T}$ and $\Psi_{nm}(\vecs^{*})=[\psi_{nm}^{1}(\vecs^{*}) \ \psi_{nm}^{2}(\vecs^{*})]^{T}$ be such that
\begin{align}
	d_{k^{*}l^{*}}(\vecr,\vecs^{*},\vecp^{*}) = d_{nm}(\Upsilon_{nm}(\vecr), \Psi_{nm}(\vecs^{*}), \vecp^{*}) \notag
\end{align}
for all $n=1,2$ and $m=1,2,\ldots,6$.  Then, the DMT in Theorem 2 is achieved by employing a sequence (in SNR) of codebooks satisfying
\begin{align}
	\| \Delta\vecu_{i}\|^2 \ &\dotgeq \ \SNR^{-\psi_{i1}^{i}(\vecs^{*}) +\epsilon} \label{faflafl}\\
     \| \Delta\vecw_{i}\|^2 \ &\dotgeq \ \SNR^{-\upsilon_{i2}^{i}(\vecr)+\psi_{i2}^{i}(\vecs^{*})+\epsilon}\notag\\
	\| \Delta\vecx_{i}\|^2 \ &\dotgeq \  \SNR^{-\upsilon_{i3}^{i}(\vecr)+\epsilon}\notag\\ 
	\lambda_{\min}(\Delta\matA_{ij}\lefto(\Delta\matA_{ij}\right)^{H}) \ &\dotgeq \ \SNR^{-\psi_{i4}^{i}(\vecs^{*}) -\upsilon_{j4}^{j}(\vecr)+\psi_{j4}^{j}(\vecs^{*})+\epsilon}\notag\\
	\lambda_{\min}(\Delta\matB_{ij}\lefto(\Delta\matB_{ij}\right)^{H}) \ &\dotgeq \ \SNR^{-\sum\limits_{k=1}^{2}\upsilon_{k5}^{k}(\vecr)+\sum\limits_{j=1}^{2}\psi_{j5}^{j}(\vecs^{*})+\epsilon} \notag\\
	\lambda_{\min}(\Delta\matC_{ij}\lefto(\Delta\matC_{ij}\right)^{H}) \ &\dotgeq \ \SNR^{-\upsilon_{i6}^{i}(\vecr) -\upsilon_{j6}^{j}(\vecr)+\psi_{j6}^{j}(\vecs^{*})+\epsilon} \notag
\end{align}
for every pair of codewords in each codebook for $i,j=1,2$, $i\neq j$, and for some$^{1}$ $\epsilon > 0$.}
\end{enumerate}
\end{theorem}
For a proof of Theorems 2 and 3, we refer  to \cite{JournalPrep}.
 \begin{remark}
Whenever $d_{HK}(\vecr) \leq d_{MU}(\vecr)$ and $\alpha=1$, the code design criteria in Theorem 3 (stated  in \eqref{xxczzxvz} and \eqref{xxczzxvz2}) are equivalent to the criteria for achieving the optimal DMT in a multiple access channel (MAC).  The existence of DMT-optimal codes for the MAC is shown in \cite{namgamal} and an explicit construction \cite{Badr08} is shown to be DMT-optimal in \cite{isit08_cgb}. For $d_{HK}(\vecr) > d_{MU}(\vecr)$, it is an open question whether there are superposition codes that satisfy \eqref{faflafl} and hence, achievability of the DMT through the fixed-power-split superposition HK scheme depends on whether this question can be resolved positively.
 \end{remark}
  
\begin{figure}[htbp]
\begin{center}
\includegraphics[scale=0.6]{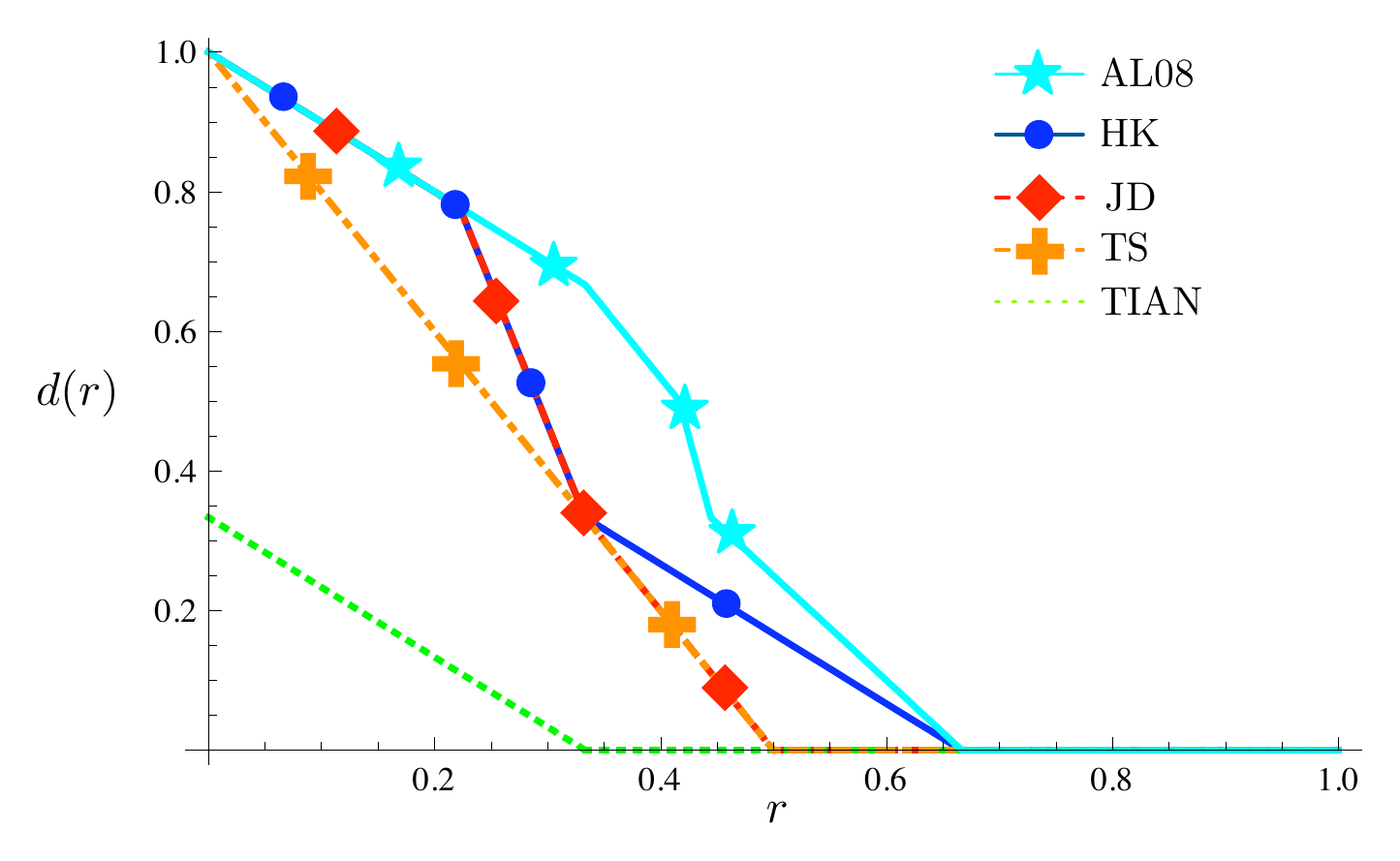}
\caption{Symmetric rate DMT for $\alpha=2/3$ and for various schemes.}
\label{div066}
\end{center}
\end{figure}

\emph{Numerical result}: For $\alpha=2/3$ and $r_{1}=r_{2}=r$,  Fig. \ref{div066} shows the DMT achieved by the fixed-power-split HK scheme (HK) in comparison to the outer bound in \cite{Akuiyibo08} (AL08), to joint decoding (JD), to treating interference as noise (TIAN), and to time-sharing (TS).   It is shown in \cite{JournalPrep} that the outer bound AL08 is loose under \emph{moderate interference}, i.e., when $2/3 \leq\alpha < 1$, and that the HK scheme is DMT-optimal in this range.  
%The IC rate region in \cite[Eq. (1)]{Akuiyibo08} is achievable by the scheme proposed by Etkin \emph{et al.} \cite{Etkin08} which, however,  makes explicit use of transmit CSI.  Hence, the result in \cite[Theorem 1]{Akuiyibo08} serves as a genie-aided outer bound on the DMT of the IC in the absence of transmit CSI, considered here. 

\bibliography{IEEEabrv,confs-jrnls,publishers,cebib}
\bibliographystyle{IEEEtran}

\end{document}